\documentclass[12pt]{article}
\usepackage{graphicx}
\usepackage{color}

\def\Title#1{\begin{center} {\Large #1 } \end{center}}
\def\Author#1{\begin{center}{ \sc #1} \end{center}}
\def\Address#1{\begin{center}{ \it #1} \end{center}}

\newenvironment{Abstract}{\begin{center}{\bf Abstract}\end{center} \bigskip \begin{quotation}  }{\end{quotation}}
\newenvironment{Presented}{\begin{quotation} \begin{center} 
             PRESENTED AT\end{center}\bigskip 
      \begin{center}\begin{large}}{\end{large}\end{center} \end{quotation}}
\def\Acknowledgements{\bigskip  \bigskip \begin{center} \begin{large}
             \bf ACKNOWLEDGEMENTS \end{large}\end{center}}

\def\beq{\begin{equation}}
\def\eeq#1{\label{#1}\end{equation}}
\def\eeqn{\end{equation}}

\def\beqa{\begin{eqnarray}}
\def\eeqa#1{\label{#1}\end{eqnarray}}
\def\eeqan{\end{eqnarray}}

\let\bar=\overbar

\def\Dslash{\not{\hbox{\kern-4pt $D$}}}
\def\dslash{\not{\hbox{\kern-2pt $\del$}}}

\def\msb{{\bar{\ssstyle M \kern -1pt S}}}

\begin{document}
\begin{titlepage}

\vfill


\Title{Measurements of the CKM angle $\phi_1/\beta$ at the $B$ Factories}
\vfill
\Author{Himansu Sahoo on behalf of the Belle and BaBar Collaborations}  
\Address{Department of Physics and Astronomy, University of Hawaii at Manoa, Honolulu, Hawaii, 96848, USA, himansu@phys.hawaii.edu}
\vfill


\begin{Abstract}
In this proceeding, we report the recent measurements of the CKM angle
$\phi_1/\beta$ using large data samples collected by the Belle and BaBar 
experiments. The experiments have collected more than 1 billion $B\overline{B}$ pairs of data sample at the $\Upsilon(4S)$ resonance using the facilities of the asymmetric-energy $e^+e^-$ colliders KEKB and PEP-II.
\end{Abstract}

\vfill

\begin{Presented}
The Ninth International Conference on\\
Flavor Physics and CP Violation\\
(FPCP 2011)\\
Maale Hachamisha, Israel,  May 23--27, 2011
\end{Presented}
\vfill

\end{titlepage}
\def\thefootnote{\fnsymbol{footnote}}
\setcounter{footnote}{0}
%


\section{Introduction}
In the standard model (SM), $CP$ violation in $B^0$
meson decays originates from an irreducible complex phase 
in the $3 \times 3$ Cabibbo-Kobayashi-Maskawa (CKM) 
mixing matrix~\cite{ckm}.
The unitarity condition of the CKM matrix gives rise to a relation
$V_{ud}V_{ub}^*+V_{cd}V_{cb}^*+V_{td}V_{tb}^* = 0$, which can be represented by a triangle in the complex plane, known as the Unitarity Triangle (UT).
The main objective of the $B$-factories is to test the SM picture of the origin of $CP$ violation by measuring 
the angles (denoted by $\phi_1$, $\phi_2$ and $\phi_3$)\footnote{BaBar uses an alternative notation $\beta$, $\alpha$ and $\gamma$ corresponding to $\phi_1$, $\phi_2$ and $\phi_3$.}
and sides of the UT using different $B$ decays. 
In this paper, we report the recent measurements concerning 
the angle $\phi_1$ ($\equiv \pi - \arg(V_{tb}^*V_{td}/V_{cb}^*V_{cd})$).

\section{Experimental Apparatus}
The measurements discussed in this paper have been obtained by the 
Belle and BaBar experiments, located at the KEKB and PEP-II asymmetry-energy $e^+e^-$ $B$ factories.
The accelerators operate at the $\Upsilon(4S)$ resonance, which is
produced with a Lorentz boost of 
0.43 at KEKB (3.5 on 8.0 GeV)~\cite{kekb}
and 0.56 at PEP-II (3.1 on 9.0 GeV)~\cite{pep2}.
The KEKB accelerator of the $B$ factory in Japan achieved the current world
record with a peak luminosity of 
$2.1 \times 10^{34}$ ${\rm cm}^{-2} {\rm s}^{-1}$. 
Both the experiments have already stopped data dating.
BaBar stopped operation in April 2008 and 
collected more than 430 $\rm{fb}^{-1}$ of data at $\Upsilon(4S)$ resonance. 
Belle stopped operation in June 2010 and 
collected more than 710 $\rm{fb}^{-1}$ of data. 
After a decade of successful operation, the $B$ factories have
a data sample of nearly $1200 \times 10^{6}$ $B\overline{B}$ pairs.
The Belle and BaBar detectors are described in 
detail elsewhere~\cite{belle,babar}.

\section{Measurements of {\boldmath $\phi_1/\beta$}}
Measurements of time-dependent CP asymmetries
in $B^0$ meson decays that proceed via the dominant CKM favored 
$b \to c \overline{c}s$
tree amplitude, such as $B^0 \to J/\psi K^0$, 
have provided a precise measurement of the angle $\phi_1$, thus 
providing a crucial test of the mechanism of $CP$ violation in the SM.
For such decays the interference between the tree amplitude 
and the amplitude from $B^0-\overline{B}{}^0$ mixing is dominated 
by the single phase $\phi_1$.
Other decay modes, which allow the measurements of $\phi_1$ are
$b \to c\bar{c}d$ transitions like 
$B^0 \to J/\psi \pi^0$, $B^0 \to D^{(*)+}D^{(*)-}$, $B^0 \to D^+ D^-$.
These modes are dominated by tree diagram, but loop may contribute.
We can also measure $\phi_1$ from pure penguin decays like
$\phi K_S^0$, $f_0 K_S^0$, $K^+K^-K^0$, $K_S^0 \pi^0$, 
$\eta' K_S^0$ and $\omega K_S^0$. 
In these decays sensitivity to new physics (NP) increases.

The $\sin 2\phi_1$ measurement from the $B$ factories is one of the main 
constraints in the global fit by CKM fitter Collaboration. 
Recently CKM fitter reported a tension ($\sim 2.8 \sigma$) between 
the measurement of ${\cal B}(B \to \tau \nu)$ and the value predicted from 
other observables excluding this measurement. So further measurements of $\sin 2\phi_1$ will help to clarify this tension.

\section{Analysis Technique}

In the $B$ meson system, the $CP$ violating asymmetry lies 
in the time-dependent
decay rates of the $B^0$ and $\overline{B}{}^0$ decays to a 
common $CP$-eigenstate ($f_{CP}$). The asymmetry can be written as:
\begin{eqnarray*}
{\cal A}_{CP}(t) &=& \frac{\Gamma[\overline{B}{}^0(t)\to f_{CP}] -  \Gamma[B^0(t)\to f_{CP}]}                  
                               {\Gamma[\overline{B}{}^0(t)\to f_{CP}] +  \Gamma[B^0(t)\to f_{CP}]}\\
                       &=& {\cal S} \sin (\Delta m_d t) + {\cal A} \cos(\Delta m_d t)
\end{eqnarray*}
where
\begin{equation}
\mathcal{S} = \frac{2 \,\rm{Im\lambda}}{|\lambda|^2+1} \hspace{2cm}
\mathcal{A} = \frac{|\lambda|^2-1}{|\lambda|^2+1}.
\end{equation}

\noindent Here $\Gamma(B^0 (\overline{B}{}^0) \to f_{CP})$ is the decay rate 
of a $B^0 (\overline{B}{}^0)$ meson decays to $f_{CP}$ 
at a proper time $t$ after the production,
$\Delta m_d$ is the mass difference 
between the two neutral $B$ mass eigenstates, $\lambda$ is a complex parameter
depending on the $B^0-\overline{B}{}^0$ mixing as well as the decay 
amplitudes of the $B$ meson decays to the $CP$ eigenstate.
The parameter ${\cal S}$ is the measure of 
mixing-induced $CP$ violation, whereas
${\cal A}$ is the measure of direct $CP$ violation\footnote{Note that BaBar uses the convention ${\cal C}$ = $-{\cal A}$.}.

In the $B$ factories, in order to measure the time-dependent 
$CP$ violation parameters, we fully
reconstruct one neutral $B$ meson into a $CP$ eigenstate.
From the remaining particles in the event, 
the vertex of the other $B$ meson is reconstructed 
and its flavor is identified.
In the decay chain 
$\Upsilon(4S)\to B^0 \overline{B}{}^0 \to f_{CP} f_{\rm tag}$, 
where one of the $B$ mesons decays at time $t_{CP}$ 
to a $CP$ eigenstate $f_{CP}$, which
is our signal mode, and the other decays at time $t_{\rm tag}$ 
to a final state $f_{\rm tag}$ that distinguishes between 
$B^0$ and $\overline{B}{}^0$, the decay
rate has a time dependence given by~\cite{carter}
\begin{eqnarray}
{\cal P}(\Delta{t})= \frac{ e^{-|\Delta{t}|/{\tau_{B^0}}} }{4\tau_{B^0}}
\biggl\{1 & + & q \cdot 
 \Bigl[ {\cal S} \sin(\Delta m_d \Delta{t}) 
 + {\cal A} \cos(\Delta m_d \Delta{t})
\Bigr] \biggr\}.
\label{eq_decay}
\end{eqnarray}
\noindent Here $\tau_{B^0}$ is the neutral $B$ lifetime,
$\Delta t = t_{CP} - t_{\rm{tag}}$,
and the $b$-flavor charge $q$ equals $+1$ ($-1$) 
when the tagging $B$ meson is identified as 
$B^0$ ($\overline{B}{}^0$).
Since the $B^0$ and $\overline{B}{}^0$ are approximately 
at rest in the $\Upsilon(4S)$ center-of-mass system,
$\Delta t$ can be determined from the displacement in $z$ 
between the $f_{CP}$ and $f_{\rm tag}$
decay vertices: $\Delta t \simeq \Delta z / (\beta \gamma c)$,
where $c$ is the speed of light.
The vertex position of the $f_{CP}$ decay is reconstructed using charged tracks
(for example, lepton tracks from $J/\psi$ in $B^0 \to J/\psi K_S^0$ decays)
and that of the $f_{\rm tag}$ decay from well-reconstructed tracks that are not
assigned to $f_{CP}$~\cite{vertexing}.
The $\Delta z$ is approximately 200 $\mu$m in Belle and 250 $\mu$m in BaBar.
We also consider the effect of detector resolution and mis-identification
of the flavor~\cite{tagging}. 
Finally, the $CP$ violation parameters are obtained from an
unbinned maximum likelihood fit to the $\Delta t$ distribution.

\section{{\boldmath $b \to c\bar{c}s$} Decay Modes}

The $b \to c\overline{c}s$ decays are known as the golden modes for
$CP$ violation measurements.
They have clean experimental signatures: many accessible modes
with relatively large branching fractions $\mathcal{O}(10^{-4})$, 
low experimental background levels and 
high reconstruction efficiencies.
These modes are dominated by
a color-suppressed $b \to c\overline{c} s$ tree diagram and the dominant
penguin diagram has the same weak phase.
The $CP$ violation comes from the $V_{td}$ element in the mixing box 
diagram, which contains the phase.
For $f_{CP}$ final states resulting from a $b \to c\bar{c}s$ transition, 
the SM predicts
$\mathcal{S} = - \xi_{CP}\sin 2 \phi_1$ and $\mathcal{A} = 0$,
where $\xi_{CP}$ is known as the $CP$ eigenvalue and have values
$+1$($-1$) for $CP$-even ($CP$-odd) final states.
The asymmetry is given as
\begin{equation}
{\cal A}_{CP} = \xi_{CP} \sin(2\phi_1) \sin(\Delta m \Delta t).
\end{equation}
We can verify this experimentally 
by measuring the number of $B^0 (\overline{B}{}^0)$ decays to $CP$ eigenstate.
Because of the high
experimental precision and low theoretical uncertainty these modes
provide a reference point in the SM. 
A non-zero value of ${\cal A}$ or any measurement of $\sin 2\phi_1$ 
that has a significant deviation indicates an evidence for NP.

Belle recently reported new measurements with its full data sample
($772 \times 10^6 B\overline{B}$ pairs) using
the modes
$B^0 \to J/\psi K^0$, $B^0 \to \psi' K_S^0$ and $B^0 \to \chi_{c1} K_S^0$.
The $J/\psi$ candidates are reconstructed from their decays to
$e^+e^-$ and $\mu^+\mu^-$, with the $K_S^0$ reconstructed from $\pi^+\pi^-$.
The $\psi'$ candidates are reconstructed from $e^+e^-$, $\mu^+\mu^-$ as well as
$J/\psi \pi^+ \pi^-$ decays. 
The $\chi_{c1}$ is reconstructed from its decays to $J/\psi \gamma$.
Belle reported nearly 15600 $CP$-odd signal events with a purity of $96\%$
and nearly 10000 $CP$-even signal events with a purity of $63\%$.
Belle observed $CP$ violation in all charmonium modes and 
the results are summarized in Table~\ref{tab:charmoniumresult}.

\begin{table}[hbtp]
\begin{center}
\begin{tabular}{c|c|c}  
\hline\hline
Decay Mode  &   $\mathcal{S}$  & $\mathcal{A}$ \\
\hline
$B^0 \to J/\psi K_S^0$ & $0.671\pm 0.029$ & $-0.014\pm0.021$\\
$B^0 \to J/\psi K_L^0$ & $-0.641\pm 0.047$ & $0.019 \pm 0.026$\\
$B^0 \to \psi' K_S^0$ & $0.739\pm0.079$ & $0.103\pm0.055$\\
$B^0 \to \chi_{c1} K_S^0$ & $0.636\pm0.117$ & $-0.023\pm0.083$\\
\hline\hline
\end{tabular}
\caption{The $CP$-violating parameters measured by Belle with golden modes 
using a data sample of $772 \times 10^6 B\overline{B}$ pairs (the errors are statistical only). Belle observed $CP$ violation in all charmonium modes.}
\label{tab:charmoniumresult}
\end{center}
\end{table}

\begin{figure}[ht]
\begin{center}
\includegraphics[width=4cm]{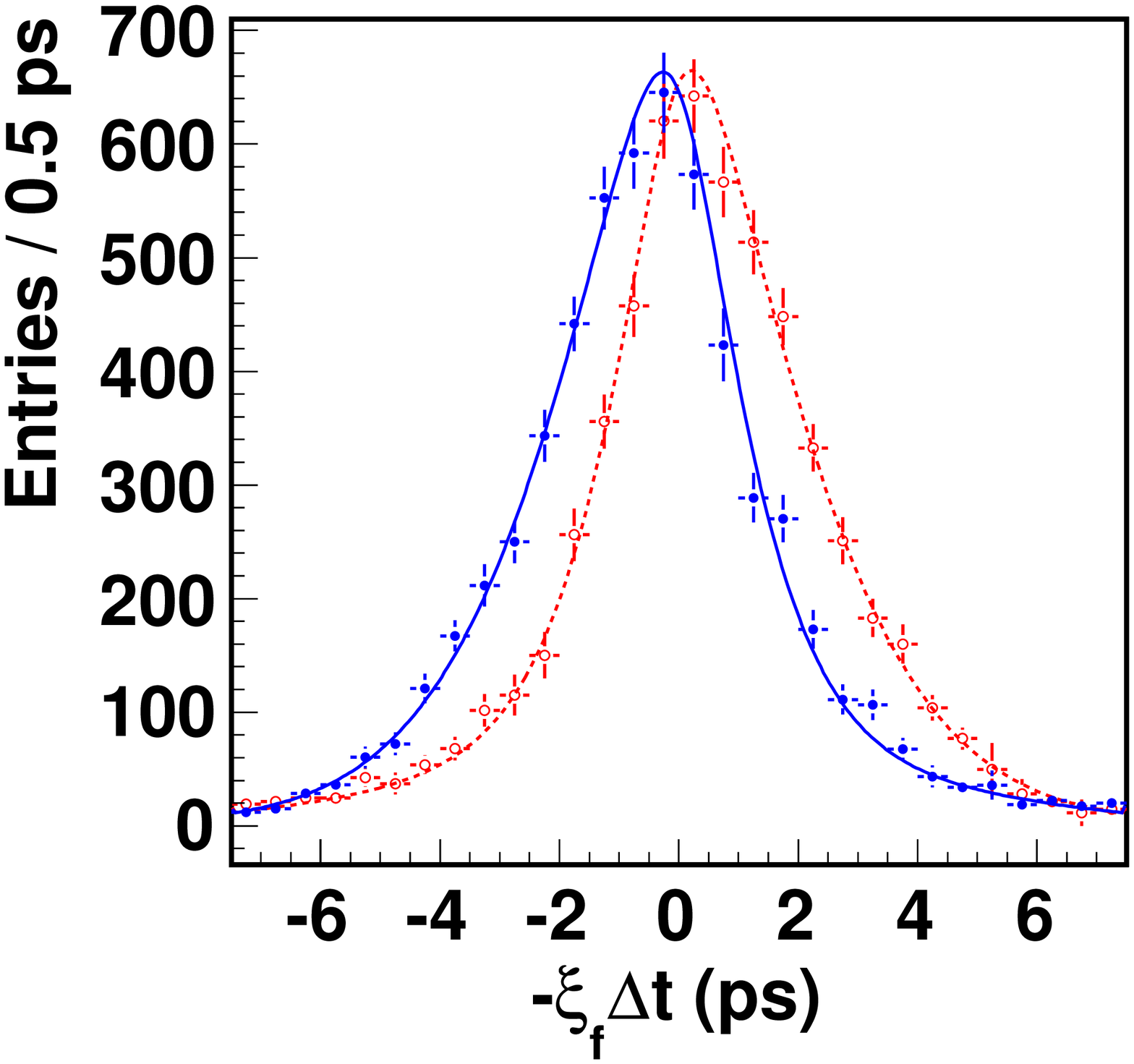}
\includegraphics[width=4cm]{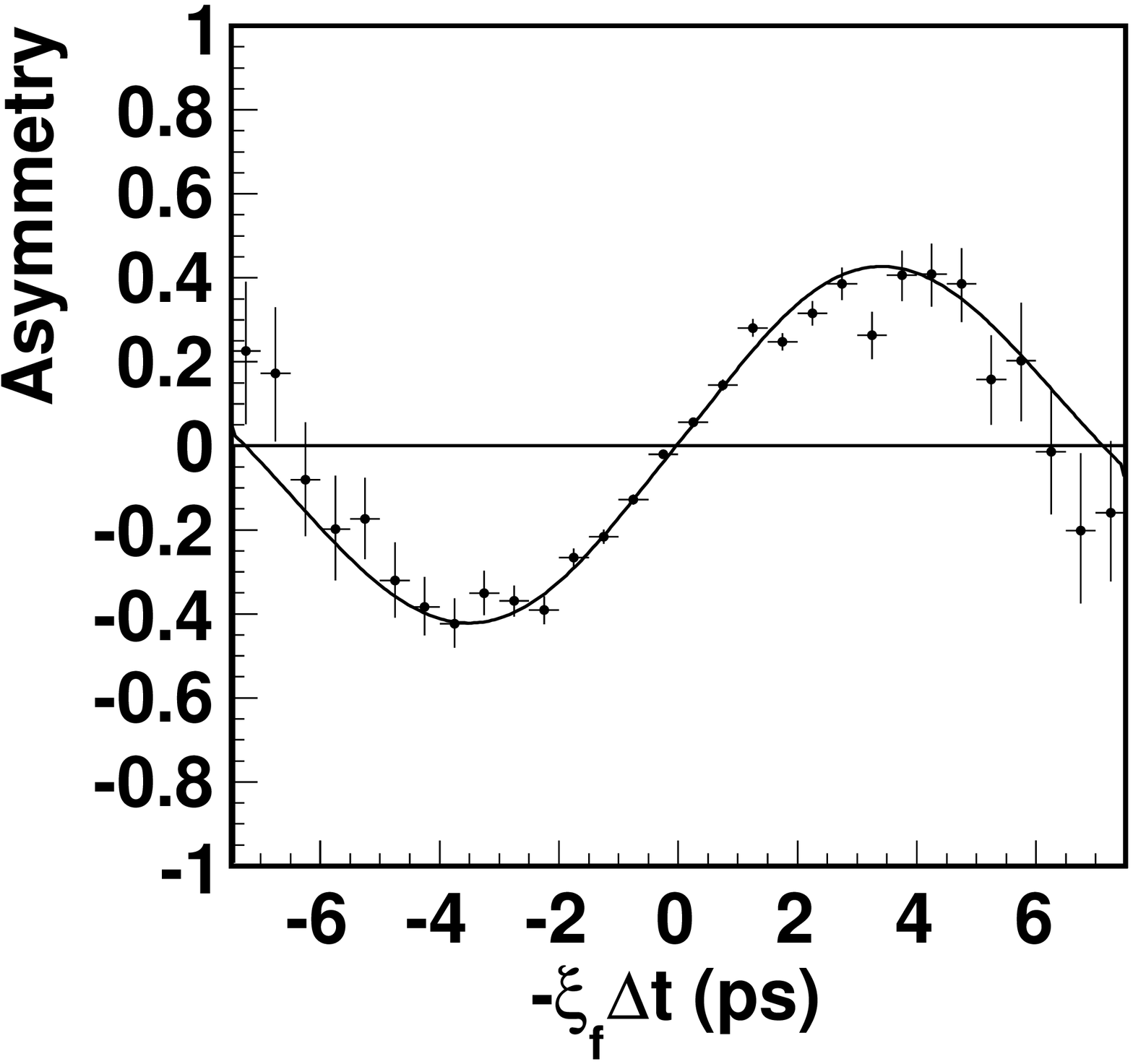}
\caption{The left plot shows the $\Delta t$ distribution for $q=+1$ (red) and $q=-1$ (blue) and right plot is the raw asymmetry. These are background-subtracted and for good-tagged events only.}
\label{fig:belledt}
\end{center}
\end{figure}

Figure~\ref{fig:belledt} shows the background-subtracted
$\Delta t$ distributions for good-tagged events only
(all charmonium modes are combined).
We define the raw asymmetry in each $\Delta t$ bin as 
$(N_{+}-N_{-})/(N_{+}+N_{-})$, where $N_{+}$ $(N_{-})$
is the number of observed candidates with $q=+1$ $(-1)$.
The systematic uncertainties are significantly improved 
compared to the previous Belle measurements~\cite{kfchen,sahoo} 
due to a better model for the resolution function (decay mode independent).
Combining all charmonium modes, Belle reported the
world's most precise measurements:
\begin{eqnarray}
\sin2\phi_1 = 0.668\pm0.023 (\rm stat) \pm 0.013 (\rm syst),
\nonumber\\
{\cal A} = 0.007\pm0.016 (\rm stat) \pm 0.013 (\rm syst).
\end{eqnarray}

BaBar reconstructed the decay modes 
$B^0 \to J/\psi K^0$, $B^0 \to \psi' K_S^0$, $B^0 \to \chi_{c1} K_S^0$, $B^0 \to \eta_c K_S^0$ and $B^0 \to J/\psi K^{*0}$. Using a data sample of
$465 \times 10^6 B\overline{B}$ pairs, BaBar reported 
nearly 8400 $CP$-odd signal events with a purity of $93\%$ and 
nearly 5800 $CP$-even signal events
with a purity of $56\%$.
Combing all charmonium modes BaBar measured
$\sin2\phi_1 = 0.687\pm0.028\pm0.012$ and 
${\cal C} = 0.024\pm0.020\pm0.016$~\cite{babar-btoccs}.

\begin{figure}[htbp]
\begin{center}
\includegraphics[width=4cm,totalheight=4cm]{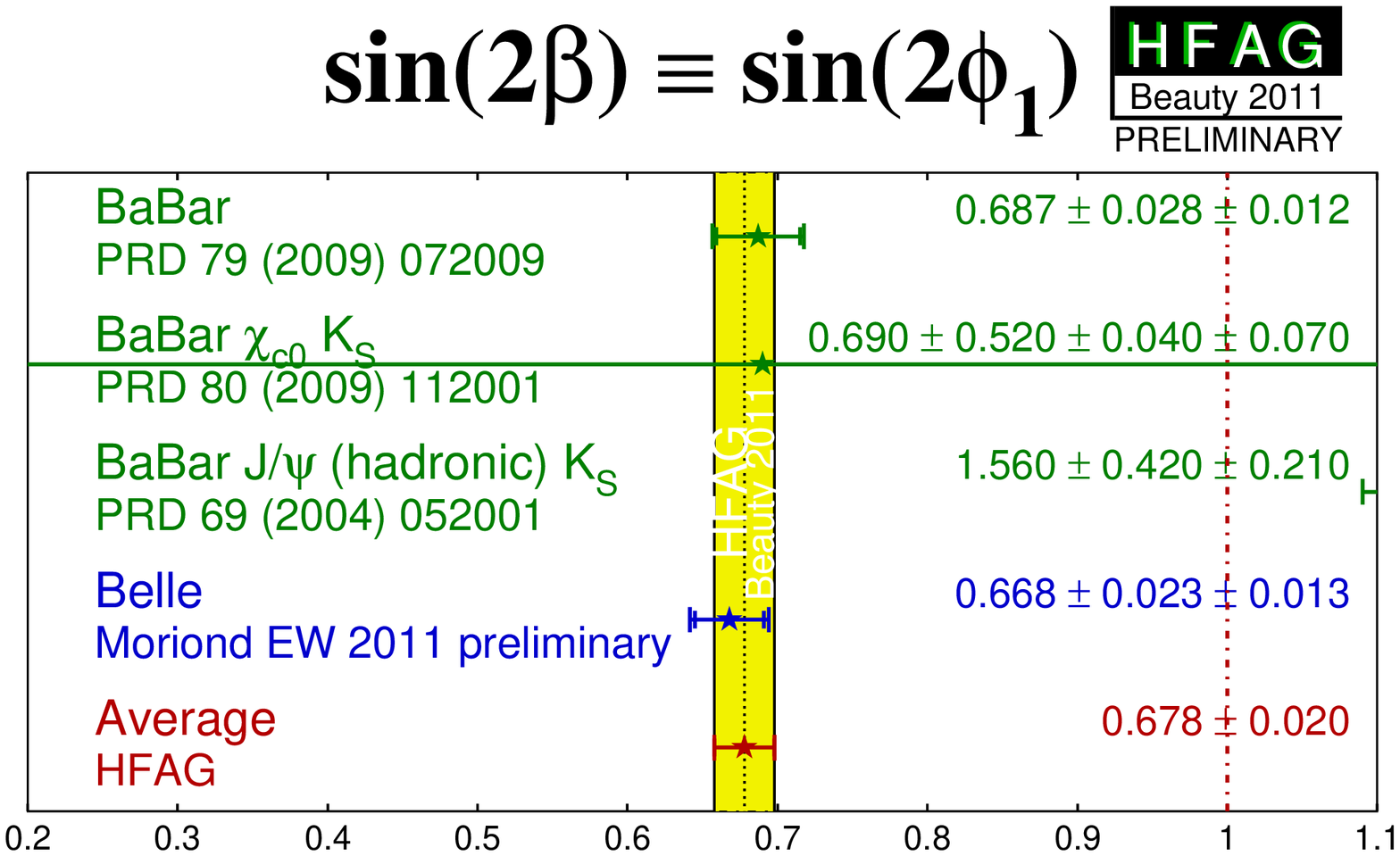}
\includegraphics[width=4cm]{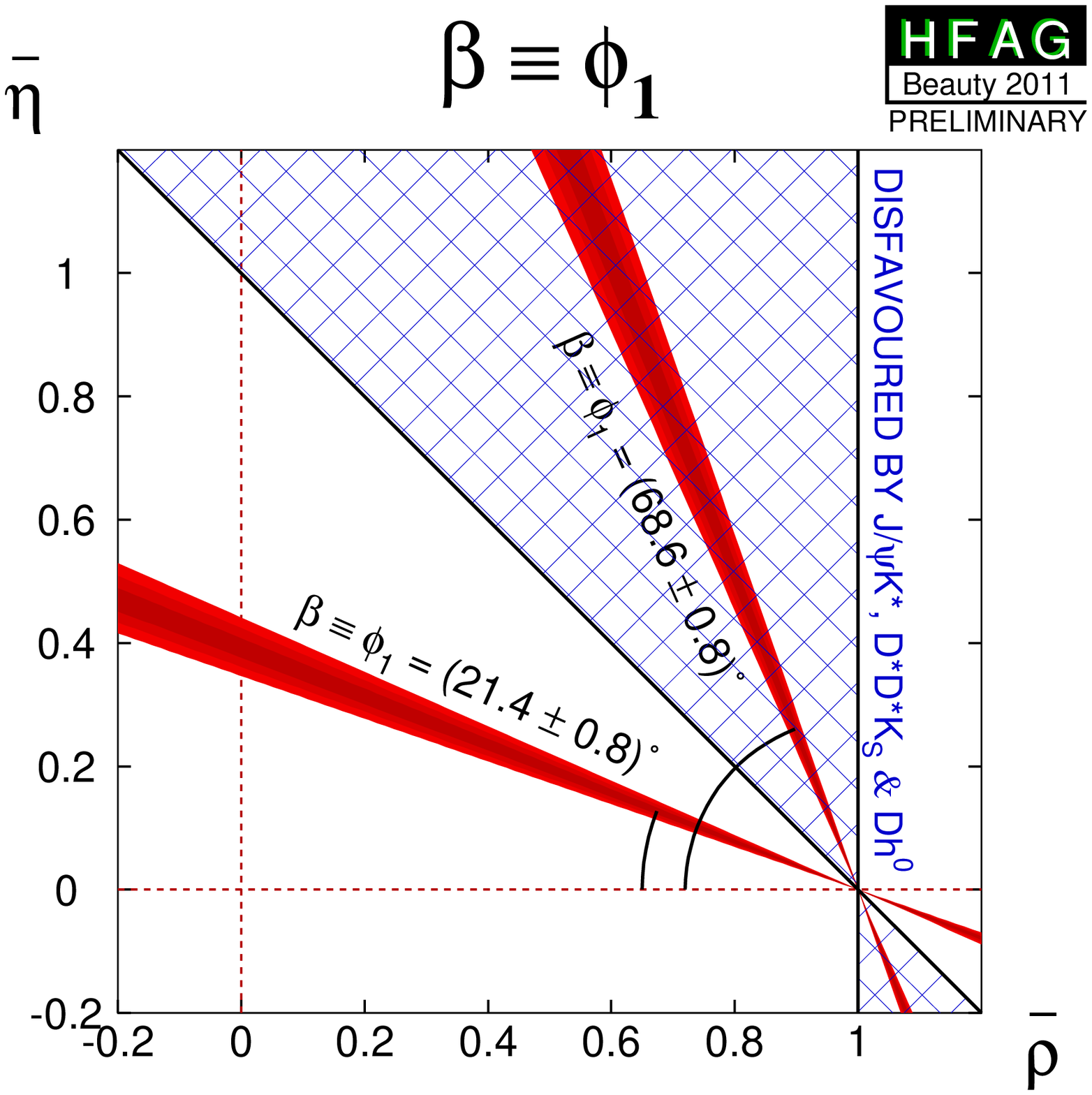}
\caption{The left plot shows the comparison between the Belle and BaBar measurements of $\sin 2\phi_1$ with $b \to c\bar{c}s$ decays and constraints on $\phi_1$ on the ($\bar{\rho},\bar{\eta}$) plane is shown in the right plot. The hatched area is excluded corresponding to the negative $\cos(2\phi_1)$ solution.}
\label{fig:averagesin2phi1}
\end{center}
\end{figure}

Combining the measurements from Belle and BaBar, the 
new world average calculated
by the Heavy Flavor Averaging Group (HFAG) is~\cite{hfag}
\begin{eqnarray}
\sin2\phi_1 (b \to c\bar{c}s) = 0.678\pm0.020,
\nonumber\\
{\cal A} (b \to c\bar{c}s) = -0.013\pm0.017.
\end{eqnarray}
Figure~\ref{fig:averagesin2phi1} summarizes the results of $\sin2\phi_1$ for
$b \to c\overline{c}s$ decays from Belle and BaBar.
The measurements of the two experiments agree very well within the statistical
uncertainties.
The experimental uncertainty on $\sin 2\phi_1$ is reduced to $3\%$ and 
thus serves as a firm reference point for the SM. 
The value of ${\cal A}$ is consistent with zero.
The new results will definitely provide 
a better constraint on the allowed region in the CKM fitter.
The measurement of $\sin 2\phi_1$ leaves a two-fold ambiguity in the value of 
$\phi_1$.
Both Belle and BaBar measured the sign of $\cos 2\beta$ to be positive at $98.3\%$ and
$86\%$ confidence levels, respectively. 
This favors the smaller value of $\phi_1$ solution.
The new measurements give the value~\cite{hfag}
\begin{eqnarray}
\phi_1 (\beta) = (21.4\pm0.8)^{\circ},
\end{eqnarray}
which is the most precise measurement with $<1^{\circ}$ error.

\section{{\boldmath $b \to c\bar{c}d$} Decay Modes}

The $B^0 \to J/\psi \pi^0$ decay takes place through a $b \to c\bar{c}d$
transition. The dominant tree diagram is Cabibbo-suppressed.
However, there is a penguin diagram of the same order that has a different
weak phase. 
So, small deviation in $\sin 2\phi_1$ from golden modes is expected in the SM. 
The BaBar result provides an evidence of $CP$ violation 
at $4\sigma$ level~\cite{jpsipi0-babar}, while the value for
Belle result is $2.4\sigma$~\cite{jpsipi0-belle}.
The decay $B^0 \to D^{*+} D^{*-}$ also goes through 
the $b \to c\bar{c}d$ transition. 
This mode requires an angular analysis to 
separate $CP$-even and $CP$-odd events.
Belle reports a statistical significance of $3.2 \sigma$ for 
direct $CP$ violation in the $B^0 \to D^+D^-$ mode~\cite{dd-fratina}.

\section{{\boldmath $b \to s q\bar{q}$} Decay Modes}

An alternative way to measure the angle $\phi_1$ is to measure
the time-dependent $CP$ asymmetries in charmless hadronic final states.
These are $b \to s$ penguin dominated decays.
Any non-SM particles, like Higgs or SUSY particles can enter
the loop. So, these are sensitive to NP.
The value of ${\cal S}$ is expected to be $\sin 2 \phi_1$ for 
a pure penguin amplitude, but can be different if there is 
an extra $CP$ phase from NP.
As a consequence, an effective $\sin2\phi_1$ value is measured.
Significant deviation from $\sin 2\phi_1$ in golden modes would indicate NP.
The deviations have been estimated in several theoretical models and are
expected to be positive. These estimates are mode and model dependent.

\begin{figure}[htbp]
\begin{center}
\includegraphics[width=5cm]{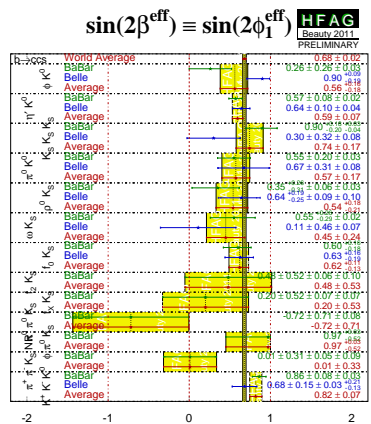}
\caption{The summary of effective $\sin2\phi_1$ measurements in $b \to s$ penguin decay modes.}
\label{fig:btoscompilation}
\end{center}
\end{figure}

Belle and BaBar have recently performed 
time-dependent Dalitz analyses in the
$B^0 \to K^+K^-K_S^0$~\cite{belle-dalitz} final state using
$657 \times 10^{6} B\overline{B}$~\cite{babar-dalitz} and
$465 \times 10^{6} B\overline{B}$ pairs, respectively.
This gives directly the value of $\phi_1$ 
(we do not need to worry about the two-fold ambiguity here).
The results are consistent with the 
SM expectation from $b \to c\bar{c}s$ decays.
The $\sin 2\phi_1^{\rm eff}$ in various $b \to s$ penguin modes 
is summarized in Fig.~\ref{fig:btoscompilation}. 
The results are consistent between Belle and BaBar
and also consistent with the SM expectation 
within the statistical  uncertainties.
It is fair to say that we need more data to see a sensitivity comparable with theoretical uncertainties.

\section{Conclusion}

In this review, we have presented the recent measurements of the CKM angle $\phi_1/\beta$ by Belle and BaBar. Thanks to the excellent performance of the two $B$ factories, which collected large 
data sample at the $\Upsilon(4S)$ resonance;
the angle $\phi_1$ has been measured with $<1^{\circ}$ precision.
The $CP$ violating parameters in $b \to c\bar{c}s$ decays are the most precise
measurements and provides a reference point for new physics searches.
The time-dependent $CP$ asymmetry in penguin dominated decays is consistent
with standard model expectations within the uncertainties of the measurement.

\Acknowledgements

I would like to thank my Belle colleagues for their valuable help in providing information regarding the measurements of $\phi_1$. 
I am also thankful to the conference organizers for their invitation to present this review.

\end{document}